\documentstyle[aps,prb,twocolumn,graphicx]{revtex}
\begin{document}

\draft
\twocolumn [\hsize\textwidth\columnwidth\hsize\csname
@twocolumnfalse\endcsname

\title
{Magnetic order and spin-waves in the quasi-1D S=1/2
antiferromagnet ${\bf BaCu_{2}Si_{2}O_{7}}$}

\author{M. Kenzelmann$^{(1)}$  \and A. Zheludev$^{(2)}$ \and  S.
Raymond$^{(3)}$ \and E. Ressouche$^{(4)}$ \and T. Masuda$^{(5,6)}$
\and P. B\"{o}ni$^{(7)}$ \and K. Kakurai$^{(8)}$ \and I.
Tsukada$^{(5,9)}$ \and  K. Uchinokura$^{(5,6)}$  and R.
Coldea$^{(10,11)}$.}

\address{(1) Oxford Physics, Clarendon Laboratory, Oxford OX1 3PU,
UK (2) Physics Department, Brookhaven National Laboratory, Upton,
NY 11973-5000, USA (3) CEA-Grenoble, DRFMC/SPSMS/MDN, 38054
GRENOBLE Cedex, FRANCE (4) DRFMC/SPSMS/MDN, CENG, 17 rue des
Martyrs, 38054 Grenoble Cedex, France (5) Department of Applied
Physics, The University of Tokyo, 6th Engineering Bldg., 7-3-1
Hongo, Bunkyo-ku, Tokyo 113-8656, Japan (6) Department of Advanced
Materials Science, The University of Tokyo, 6th Engineering Bldg.,
7-3-1 Hongo, Bunkyo-ku, Tokyo 113-8656, Japan (7) Laboratory for
Neutron Scattering ETH \& PSI, CH-5232, Villigen PSI, Switzerland
(8) Neutron Scattering Laboratory, Institute of Solid State
Physics, The University of Tokyo, Tokai, Ibaraki 319-1106, Japan
(9) Central Research Institute of Electric Power Industry, 2-11-1,
Iwato kita, Komae-shi, Tokyo 201-8511, Japan (10) Oak Ridge
National Laboratory, Solid State Division, Oak Ridge, TN 37831,
USA (11) ISIS Facility, Rutherford Appleton Laboratory, Oxon OX11 0QX, UK}%

\date{\today}
\maketitle
\begin{abstract}
Elastic and inelastic neutron scattering were used to study the
ordered phase of the quasi-one-dimensional spin-1/2
antiferromagnet ${\rm BaCu_{2}Si_{2}O_{7}}$. The previously
proposed model for the low-temperature magnetic structure was
confirmed. Spin wave dispersion along several reciprocal-space
directions was measured and inter-chain, as well as in-chain
exchange constants were determined. A small gap in the spin wave
spectrum was observed and attributed to magnetic anisotropy
effects. The results are discussed in comparison with existing
theories for weakly-coupled quantum spin chain antiferromagnets.
\end{abstract}

 ]

\newpage

\section{Introduction}
The investigation of low-dimensional magnets has been a very
active research field for the last two decades. First conceived as
simple spin models, low-dimensional magnets revealed a far richer
physical behaviour than their more conventional three-dimensional
(3D) counterparts, due to the importance of quantum fluctuations.
Particularly interesting is the one-dimensional (1D) Heisenberg
antiferromagnetic model (1D HAF), for which the Hamiltonian  is
written as
\begin{equation}
H = J \sum_{\rm i}^{\rm chain} \bbox{S}_{\rm i} \cdot
\bbox{S}_{\rm i+1} \, , \label{Hamiltonian}
\end{equation}
$J$ being the exchange coupling constant. The ground state of the
1D S=1/2 HAF is a spin-singlet and can be characterized as
``marginal spin liquid'': while long-range order is absent,
spatial spin correlations decay according to a power-law, and thus
represent quasi-long-range order.\cite{H_A_Bethe} The excitation
spectrum is isotropic and gapless, and is described in terms of
$S=1/2$ elementary excitations known as
``spinons''.\cite{Faddeev_Takhatajan} Physical excitations are
composed of pairs of spinons, which gives rise to a
2-spinon-continuum.\cite{G_Muller} This behavior is in stark
contrast with that of 3D spin systems, that have long-range order
in the ground state (``spin solid''), and where the spectrum is
dominated by single-particle $S=1$ excitations (spin waves). The
pure one-dimensional HAF is of course a physical abstraction since
spin chains in real materials are always at least weakly coupled.
To make a virtue of necessity, weakly coupled antiferromagnetic
chains offer the opportunity to study the cross-over from the
quantum spin dynamics in a single chain to semi-classical spin
waves in 3D systems.\par

\begin{figure}
\begin{center}
  \includegraphics[height=5.5cm,bbllx=64,bblly=265,bburx=560,
  bbury=570,clip=]{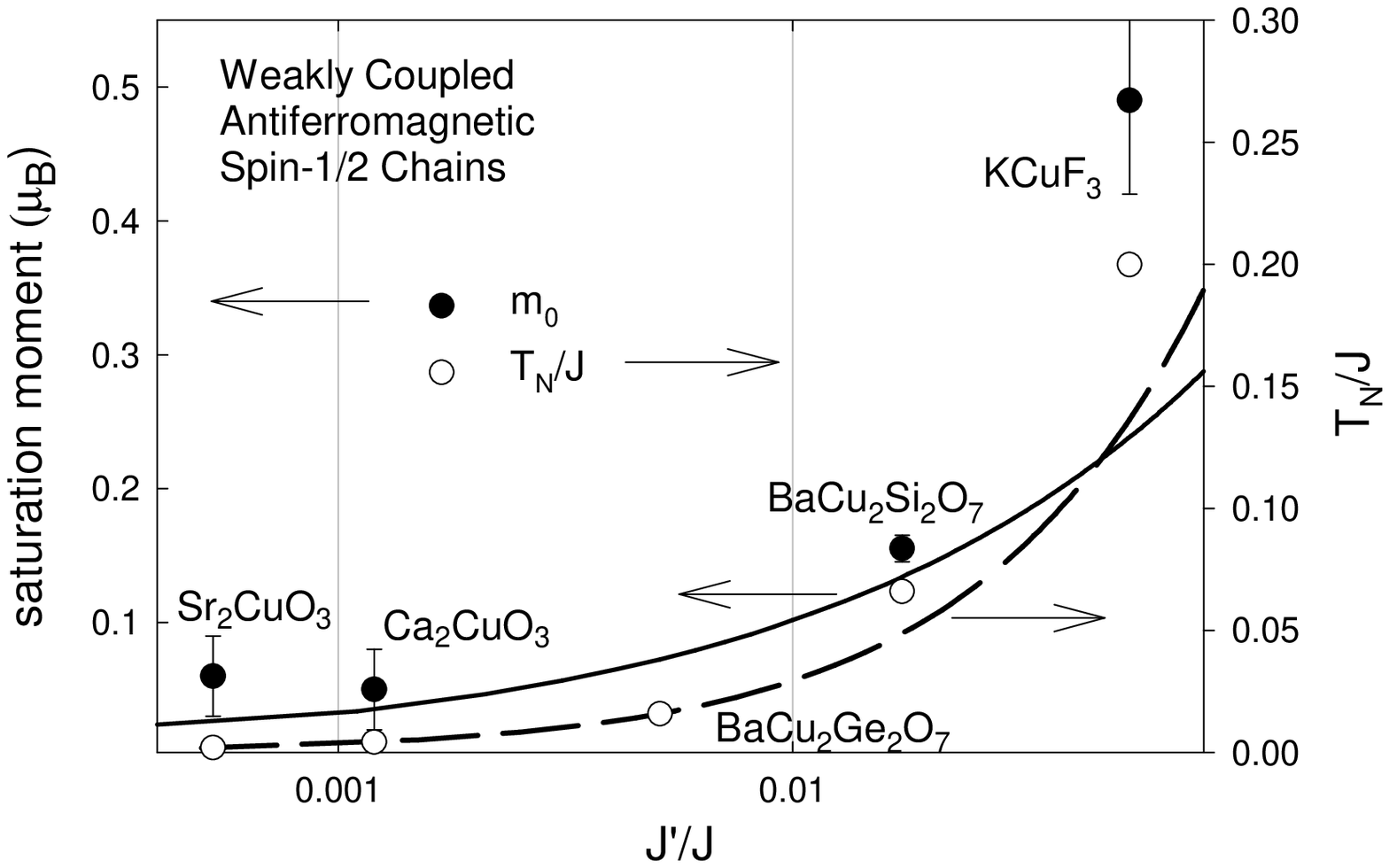}
  \vspace{0.3cm}
  \caption{The ordered moment and $T_{N}/J$ for the known
  weakly-coupled antiferromagnetic Heisenberg spin-1/2 chain
  compounds as a function of $|J'|/J$. For ${\rm
  Sr_{2}CuO_{3}}$~\protect\cite{Kojima_Sr2CuO3_Ca2CuO3},
  ${\rm Ca_{2}CuO_{3}}$~\protect\cite{Yamada_Ca2CuO3}
  and ${\rm BaCu_{2}Ge_{2}O_{7}}$~\protect\cite{Tsukada_BaCu2Si2O7,Tsukada_BaCu2Ge2O7},
  $|J'|$ was estimated using the predictions of the
  chain- mean field model,\protect\cite{Schulz96} while actual measured
  values were used for $T_{N}$. For ${\rm
  KCuF_{3}}$~\protect\cite{Satija_KCuF3,Hutchings} $|J'|$ was
  deduced from spin wave bandwidth perpendicular to the chain
  axis. The lines are predictions of  the chain mean-field model
  \protect\cite{Schulz96}.}
  \label{comparison}
\end{center}
\end{figure}

${\rm BaCu_{2}Si_{2}O_{7}}$ was recently recognized as an almost
ideal model system \cite{Tsukada_BaCu2Si2O7} for experimental
investigation of this dimensional cross-over. The silicate
crystallizes in an orthorhombic crystal structure, space group
\textit{Pnma}, and the lattice constants are $a = 8.862(2)\,$\AA,
$b = 13.178(1)\,$\AA \, and $c =
6.897(1)\,$\AA.\cite{Oliveira_Ba2CuSi2O7} The ${\rm
Cu^{2+}}$-spins are subject to a strong super-exchange interaction
via the ${\rm O^{2-}}$-ions, producing weakly-coupled
antiferromagnetic chains that run along the crystallographic
$c$-axis.\cite{Satija_KCuF3} Correspondingly, the magnetic
susceptibility shows a broad maximum at around $150\;\mathrm{K}$
\cite{Tsukada_BaCu2Si2O7} and its temperature dependence follows
the theoretical Bonner-Fisher curve \cite{Bonner_Fisher} with an
intrachain coupling $J=24.1\;\mathrm{meV}$. The nearest-neighbor
Cu-Cu distance within the chains is equal to $c/2$ (for a
schematic view of the crystal structure see Fig.~1 in Tsukada
\textit{et al.}\cite{Tsukada_BaCu2Si2O7}). Weak interactions
between the chains result in long-range antiferromagnetic ordering
at $T_{N }=9.2\;\mathrm{K}$, as observed by bulk susceptibility
and specific heat measurements.\cite{Tsukada_BaCu2Si2O7}
Preliminary neutron diffraction studies provided an estimate for
the ordered magnetic moment at low temperatures: $m_{\rm
0}=0.16\mu_{B}$. \cite{Tsukada_BaCu2Si2O7} In the ordered state
the spins are aligned along the crystallographic $c$-axis (chain
axis). Early inelastic work suggested that the ratio of
inter-chain to in-chain interactions $J'/J$ in ${\rm
BaCu_{2}Si_{2}O_{7}}$ is intermediate compared to that in such
well-known $S=1/2$ systems as ${\rm KCuF_{3}}$ (a more 3D-like
material) and ${\rm Sr_{2}CuO_{3}}$ (an almost perfect 1D
compound), as shown in Fig.~\ref{comparison}.\par

In a recent short paper we reported studies of the interplay
between single-particle and continuum dynamics in  ${\rm
BaCu_{2}Si_{2}O_{7}}$.\cite{Zheludev_BaCu2Si2O7} For a better
quantitative understanding of this behavior a detailed knowledge
of both in-chain and inter-chain interactions in this system is
required. The present article is a detailed report on neutron
scattering characterization of this material. We focus on the 3D
aspect, namely long-range magnetic order and spin wave-like
excitations. From measurements of the spin wave dispersion along
different reciprocal-space directions we determine all the
relevant parameters of the system, such as the strengths and
geometry of inter-chain coupling and magnetic anisotropy.

\section{Experimental}
Single crystals were grown using the floating-zone technique from
a sintered polycrystalline rod. The crystals were carefully
checked for twinning because the lattice constants along the $a$-
and $c$-axis are very similar. The diffraction measurements were
performed with a small sample with dimensions $3\;\mathrm{mm}^3$
to exclude sizable extinction effects. The samples used in the
inelastic scattering experiments were cylindrical in shape,
$5\;\mathrm{mm}$ in diameter and $50\;\mathrm{mm}$ in height. Up
to three single crystals were co-aligned for the different
experiments.\par

The neutron scattering experiments were performed on five
different instruments. The two-axis D23 diffractometer at the
Institut Laue-Langevin (ILL), Grenoble, France, was used to
investigate the ordered magnetic structure at low temperatures
(Setup I). D23 is a double-monochromator thermal-neutron
diffractometer with a lifting detector arm. The experiment was
performed with graphite monochromators and an incident energy
$E_{i}=14.7\;\mathrm{meV}$.
\par

Spin wave dispersion perpendicular to the chain axis was measured
using two cold-neutron triple-axis spectrometers: the TASP
spectrometer at Paul Scherrer Institut (PSI), Villigen,
Switzerland (setup II), and the IN14 spectrometer at ILL (Setup
III). The zone-boundary energy was measured using the thermal
neutron triple-axis spectrometer IN22 at ILL (Setup IV).
Preliminary measurements were also performed at the National
Institute of Standards and Technology Center for Neutron Research
(NCNR).\par

The experiment on TASP was performed with final neutron energy
fixed at $E_{f}=8\;\mathrm{meV}$. PG(002) reflections were used
for monochromator and analyzer. The sample was mounted with its
$(h,h,l)$ and $(h,0,l)$ crystallographic plane in the horizontal
scattering plane. The supermirror neutron guides gave a
source-to-monochromator collimation of 70' both horizontally and
vertically. The measurements in the $(h,0,l)$-plane were performed
using 80'-80' collimators before and after the sample and distance
collimation after the analyzer. For the measurements in the
$(h,h,l)$-plane, the use of $2\;\mathrm{cm}$ wide slits cut the
beam divergence before and after the sample down to effective
values of 47'-60'. The energy resolution was $0.38\;\mathrm{meV}$
and $0.6\;\mathrm{meV}$ for the measurements in the $(h,h,l)$ and
$(h,0,l)$ crystallographic plane, respectively, as determined from
the full width at half maximum (FWHM) of the quasi-elastic
peak.\par

For the measurements using IN14 and IN22, PG(002) reflections were
used for monochromator and analyzer. The sample was mounted with
its $(0,k,l)$ crystallographic plane in the horizontal scattering
plane. Final neutron energy was fixed at $E_{f}=3\;\mathrm{meV}$
and $E_{f}=30.5\;\mathrm{meV}$, for the IN14 and IN22,
respectively, with neutron guides defining the
source-to-monochromator collimations of 30' and 60', respectively.
40'-40' collimators were used in both setups before and after the
sample, and the use of slits cut the effective collimation of the
IN22 setup down to 27'-40'. The energy resolution (FWHM) was
$0.26\;\mathrm{meV}$ and $1\;\mathrm{meV}$ for the IN14 and IN22
measurements, respectively. Be- or pyrolytic graphite filters were
used after the sample to eliminate higher-order beam
contamination.\par

The time-of-flight chopper spectrometer MARI at the ISIS Facility,
Rutherford Appleton Laboratory, Oxon, United Kingdom, was used for
the measurement of the magnetic excitations in a wide range of
wave-vector transfers along the chain (Setup V). The measurements
were performed using a chopper-monochromated incident energy
$E_{i}=30\;\mathrm{meV}$ allowing the measurement of the
excitation spectrum up to $25\;\mathrm{meV}$. The energy
resolution was about $0.35\;\mathrm{meV}$, as determined from the
FWHM of the quasi-elastic peak. The resolution of the wave-vector
transfer was given by the size of the sample, the detector and
their respective distance, and it was typically $0.05$ r.l.u.
along the chain-axis.

\section{Magnetic Structure}\label{section_magn_structure}
To determine the spin structure in the ordered phase, $48$
magnetic Bragg intensities were measured at $T=1.5\;\mathrm{K}$
using Setup I. The main complication in the experiment was that
all magnetic reflections appear on top of strong nuclear peaks.
The following measurement procedure was implemented. For each
reflection the nuclear intensity was measured in a rocking curve
at $T=10~{\rm K}~>T_{N}$. The peak intensity was then measured
with high statistics both above and below $T_{N}$, and the
difference was attributed to magnetic scattering. The integrated
magnetic intensity was estimated from the measured intensity ratio
and the integrated nuclear intensity. The magnetic structure was
found to be totally consistent with that suggested in Tsukada
\textit{et al.}.\cite{Tsukada_BaCu2Si2O7} The ordered moment at
low temperature $m_{0}=0.15(1)\mu_{B}$ is parallel to the
crystallographic $c$-axis. Relative nearest-neighbor spin
alignment is ferromagnetic along the $a$-axis and
antiferromagnetic along the $b$- and $c$-axes, respectively.\par

\begin{figure}[phtb]
\begin{center}
  \includegraphics[height=6cm,bbllx=55,bblly=260,bburx=490,
  bbury=580,clip=]{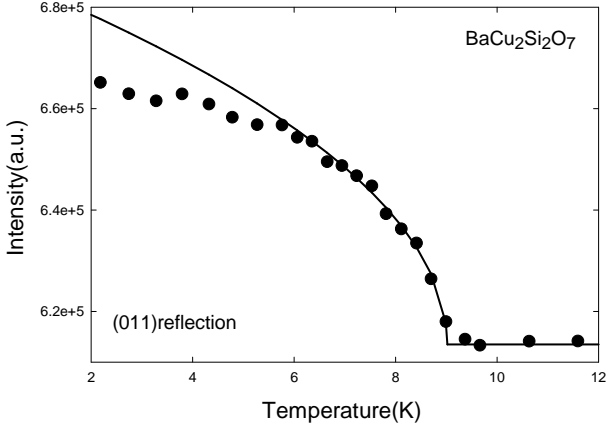}
  \vspace{0.3cm}
  \caption{Temperature dependence of the $(0\;1\;1)$-reflection
  between $1.5$ and $15\;\mathrm{K}$. The scattering intensity above
  $T_{\rm N}=9.02\;\mathrm{K}$ comes from the allowed nuclear Bragg
  reflection at $(0\;1\;1)$. Below $T_{\rm N}$, the intensity
  increases with decreasing temperature. The solid line is a fit
  to the data with the critical exponent $\beta=0.25$ as explained
  in the text.}
  \label{magnetic_structure}
\end{center}
\end{figure}

The temperature dependence of a few magnetic reflections was
measured in the range $T=1.5$-$15\;\mathrm{K}$ as shown in
Fig.~\ref{magnetic_structure} for the $(0\;1\;1)$-peak. Below the
ordering temperature $T_{N}$, the Bragg intensity increases with
decreasing $T$. The measured peak intensity $I$ is a sum of
nuclear and magnetic contributions, the latter being proportional
to the square of the sublattice magnetization $m$. The data were
fit to a power-law form:
\begin{equation}
          I(T) = I_0 \left(\frac{T_{N}-T}{T_{N}}\right)^{2\beta}
          + N \;\; {\rm for} \;\; T < T_{N}\, ,
          \label{Eq_magnetization}
\end{equation}where $I_0$ is the magnetic intensity at
$T=0\;\mathrm{K}$, $\beta$ is the order-parameter critical
exponent, and $N$ is the (nuclear) scattering intensity assumed to
be $T$-independent. A good fit is obtained with $\beta=0.25(5)$
and $T_{N}=9.0(0.05)\;\mathrm{K}$. The value of $\beta$ is clearly
below the critical exponent for a 3D antiferromagnet, and is in
good agreement with the critical indexes measured in the quasi-1D
materials ${\rm KCuF_{3}}$ \cite{D_Tennant_2} and ${\rm
Sr_{2}CuO_{3}}$\cite{Kojima_Sr2CuO3_Ca2CuO3}.

\section{Dispersion of magnetic excitations}
\subsection{Dispersion along $[h,0,1]$ and $[h,h,1]$.}

\begin{figure}
\begin{center}
  \includegraphics[height=8cm,bbllx=25,bblly=35,bburx=560,
  bbury=590,clip=]{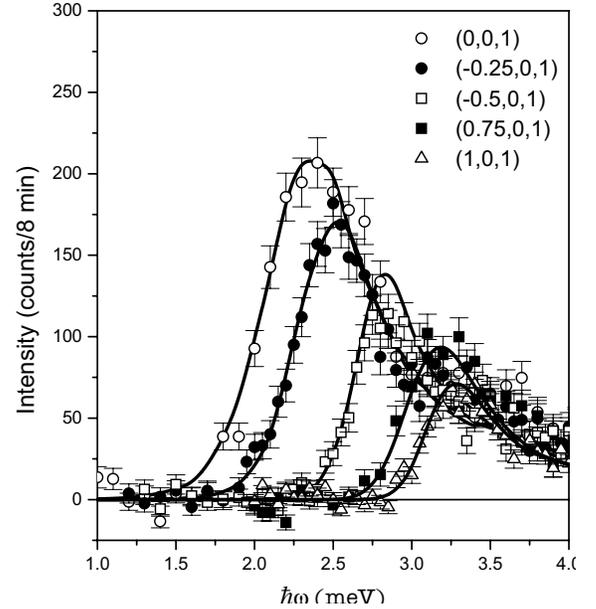}
  \vspace{0.3cm}
  \caption{Constant-$Q$ scans measured at $T=1.5\;\mathrm{K}$ for three
  different wave-vector transfers on the $[h,0,1]$
  reciprocal-space rod (Setup II). The background has been
  subtracted as described in the text. The solid lines represent
  global fits to the data using a model SMA cross section function.}
  \label{TASP-scans-h01}
\end{center}
\end{figure}

\begin{figure}
\begin{center}
  \includegraphics[height=8cm,bbllx=25,bblly=35,bburx=560,bbury=590,clip=]
  {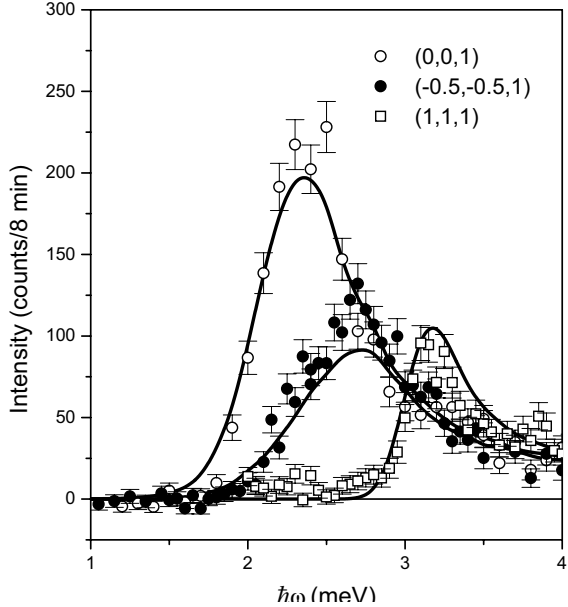}
  \vspace{0.3cm}
  \caption{Constant-$Q$ scans collected at $T=1.5\;\mathrm{K}$ for three
  different wave-vector transfers on the $[h,h,1]$ reciprocal-space rod
  (Setup II). Background and lines as in Fig.~\protect\ref{TASP-scans-h01}.}
  \label{TASP-scans-hh1}
\end{center}
\end{figure}

Spin wave dispersion at the 1D AF zone-center $q_\|=\pi$ was
measured along the $[1,0,0]$ and $[1,1,0]$ reciprocal-space
directions at $T=1.5\;\mathrm{K}$ using Setup II. Typical
background-subtracted constant-$Q$ scans are shown in
Figs.~\ref{TASP-scans-h01} and \ref{TASP-scans-hh1}. The
background was measured at $\bbox{Q}=(h,0,1.25)$ $(h$=$0-1)$ and
$\bbox{Q}=(h,h,1.25)$ $(h$=$0-1)$, where no magnetic scattering is
expected in the relevant energy transfer range, due to the steep
dispersion along the  chain axis. All measured scans show typical
spin wave peaks with a sharp onset on the low-energy side, and an
extended high-energy resolution tail.\par

The data were analyzed using a model cross-section  written in the
single-mode approximation (SMA), as will be discussed in detail in
section \ref{discussions}. The cross section was numerically
convoluted with the spectrometer resolution function, calculated
in the Cooper-Nathans approximation.\cite{Cooper_Nathans} Global
fits to all data, as well as fits to individual scans were
performed in the energy transfer range $0$-$4.5\;\mathrm{meV}$. As
shown in Figs.~\ref{TASP-scans-h01} and \ref{TASP-scans-hh1}, the
model used gives an accurate description of the experimentally
observed peak shapes. The transverse spin wave dispersion relation
deduced from the global fit is shown in Fig.~\ref{dispersion}
(solid lines), where symbols represent excitation energies
obtained in fits to individual scans.\par

\subsection{Dispersion along $[0,k,1]$}

\begin{figure}
\begin{center}
  \includegraphics[height=7cm,bbllx=40,bblly=360,bburx=510,
  bbury=760,clip=]{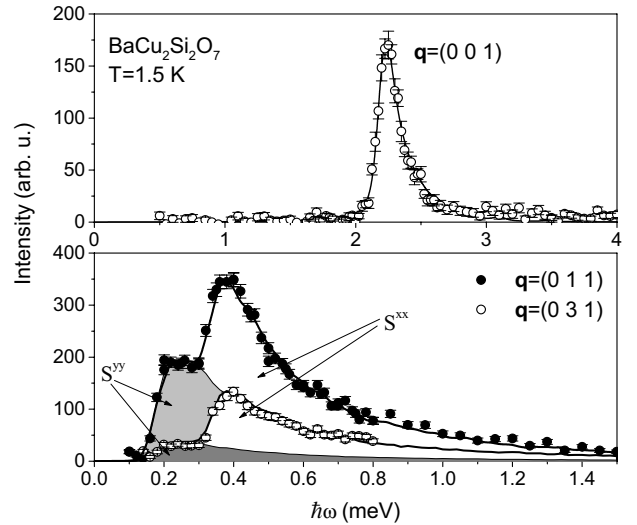}
  \vspace{0.3cm}
  \caption{Constant-$Q$ scans collected at $T=1.5\;\mathrm{K}$ for
  $\bbox{Q}=(0,0,1)$, $\bbox{Q}=(0,1,1)$ and $\bbox{Q}=(0,3,1)$ (Setup III).
  The solid line is a global fit to the data as described in the text.
  Shaded areas indicate partial contributions of two spin wave branches
  with different polarization.}
  \label{mode-splitting}
\end{center}
\end{figure}

Dispersion along the $[0,k,1]$-direction was measured using the
high-resolution cold-neutron Setup III at $T=1.5\;\mathrm{K}$.
Constant-$Q$ scans were collected at wave vectors with $k$ in the
range $0 \leq k \leq 3$. Some of these data were presented
elsewhere.\cite{Zheludev_BaCu2Si2O7} Scans at the 3D AF
zone-centers $(0,1,1)$ and $(0,3,1)$ (Fig.~\ref{mode-splitting})
reveal the presence of a small gap in the spin wave spectrum. In
fact, there appear to be two separate modes with slightly
different gap energies. These two branches correspond to different
polarizations. The relative intensity of the lower-energy
excitation at $\bbox{Q}=(0,3,1)$ is clearly smaller than at
$\bbox{Q}=(0,1,1)$. Due to the intrinsic polarization dependence
of the magnetic neutron cross section, the observed behavior is
consistent with the higher-energy mode being polarized along the
$a$-axis of the crystal, and the lower-energy excitation polarized
along $b$. This consideration is made quantitative by a global fit
to all measured scans using an SMA cross section with built-in
neutron polarization factors for the two branches (solid lines in
Fig.~\ref{mode-splitting}).\par

\begin{figure}
\begin{center}
  \includegraphics[height=6cm,bbllx=70,bblly=270,bburx=515,
  bbury=570,angle=0,clip=]{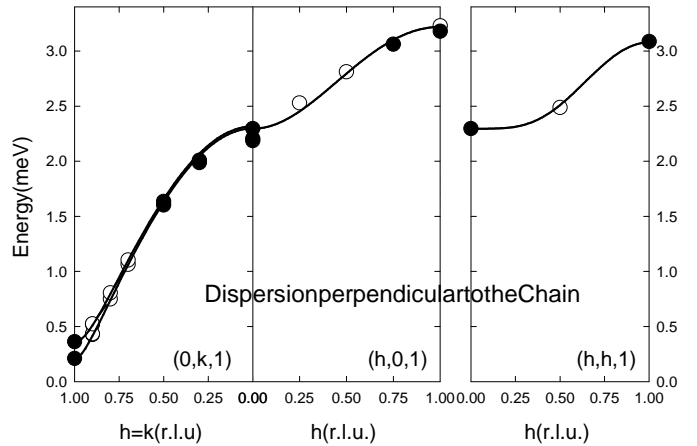}
  \vspace{0.3cm}
  \caption{Dispersion of the spin-wave excitations at $T=1.5\;\mathrm{K}$
  along three different crystallographic directions perpendicular to the
  chain axis. The open and closed circles show the excitation
  energies extracted from fits to individual scans. The open
  circle indicates that the measurements were actually performed at
  $(0,1-k,1)$, $(-h,0,1)$ and $(-h,-h,1)$, respectively.
  Error bars are smaller than the symbol size. The solid line is the
  dispersion determined in a global fit to the data using
  Eq.~\protect\ref{Eq_dispersion}.}
  \label{dispersion}
\end{center}
\end{figure}

The spin wave dispersion relation along the $[0,k,l]$
reciprocal-space direction is shown in Fig.~\ref{dispersion}
(solid lines). Here symbols represent fits to individual
scans.\par

\subsection{Dispersion along the chain axis}
\begin{figure}
\begin{center}
  \includegraphics[height=6cm,bbllx=125,bblly=260,bburx=500,
  bbury=550,clip=]{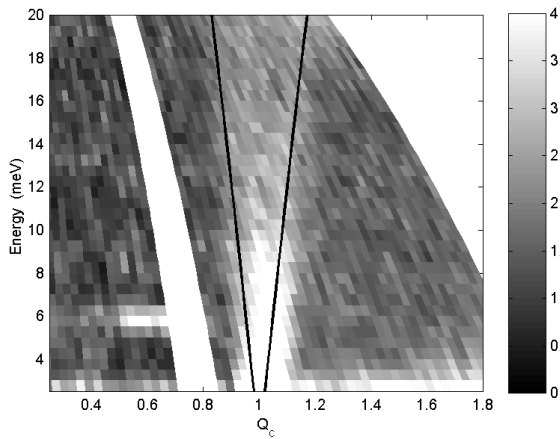}
  \vspace{0.3cm}
  \caption{Neutron scattering intensity measured at $8.5\;\mathrm{K}$
  using MARI as a function of wave-vector transfer along the chain-axis
  and energy transfer. The color indicates the measured intensity
  according to the black and white colorbar on the side. The empty
  spaces are due to the detector arrangement which is not continuous
  in the $2\Theta$ scattering angle. The black line is the lower bound
  of the two spinon continuum for $J=24.1\;\mathrm{meV}$.}
  \label{MARI-shot}
\end{center}
\end{figure}

The dispersion of magnetic excitations along the chains was
previously determined in constant-energy scans up to
$16.5\;\mathrm{meV}$ energy transfer. \cite{Tsukada_BaCu2Si2O7} As
part of the present work we performed additional measurements of
this type using the time-of-flight technique (Setup V). The
incident neutron energy at the MARI spectrometer was set to
$30\;\mathrm{meV}$ and the chain axis was aligned nearly
perpendicular to the incoming beam direction. The scattered
neutrons were counted by a large array of detectors, arranged in a
half circle with radius $4\;\mathrm{m}$ vertically below the
sample. The energy transfer was determined from the detection
time. This allowed a simultaneous measurement of the neutron
scattering cross section on a two-dimensional surface in
$(\bbox{Q},\omega)$-space. The projection of the measurement
surface onto the $(Q_{\|},\omega)$-plane ($Q_{\|}$ being the wave
vector transfer along the chain axis) is shown in
Fig.~\ref{MARI-shot} for $T$=$8.5\;\mathrm{K}$. One clearly sees a
``fountain'' of magnetic scattering emanating from the
$Q_{\|}=2\pi/c$ 1D AF zone-center. We have previously demonstrated
that in this energy range both spin waves and continuum
contributions to the magnetic dynamic structure factor are
important.\cite{Zheludev_BaCu2Si2O7} However, in the present
experiment separating the two does not appear possible due to a
poor wave vector resolution (typically $0.05$ r.l.u. along the
chains). The observed dispersion of magnetic excitations is
consistent with the previous estimates of $v_s$. This is
illustrated by the solid line in Fig.~\ref{MARI-shot} that shows
the calculated Des Cloizeaux -Pearson SMA dispersion relation for
a 1D $S=1/2$ Heisenberg antiferromagnet,\cite{Cloizeaux_Pearson}
or the lower bound of the 2-spinon excitation
continuum,\cite{G_Muller} assuming $v_{s}
=130.5\;\mathrm{meV\cdot}$\AA.

\subsection{Zone-boundary energy}
\begin{figure}
\begin{center}
  \includegraphics[height=6cm,bbllx=64,bblly=265,bburx=480,
  bbury=570,clip=]{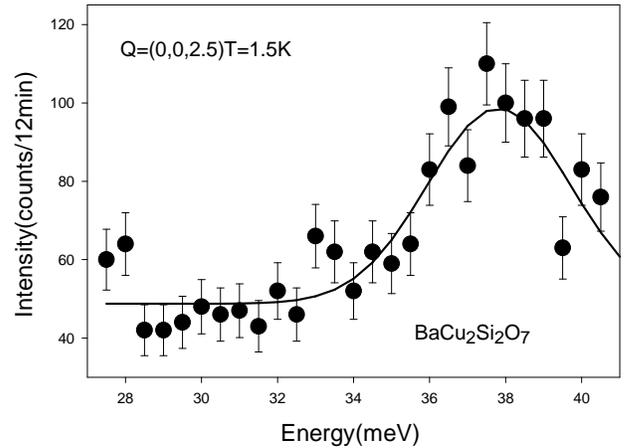}
  \vspace{0.3cm}
  \caption{Constant-$Q$ scan collected at $T=1.5$~K at the 1D
  antiferromagnetic zone boundary $\bbox{Q}=(0,0,2.5)$. The solid
  line is a Gaussian fit to the data.}
  \label{zone-boundary}
\end{center}
\end{figure}

To verify that only nearest-neighbor in-chain interactions are
relevant, the measured value of $v_s$ was compared to the
excitation energy at the 1D AF zone-boundary. The latter was
measured at $T=1.5\;\mathrm{K}$ in a constant-$Q$ scan at the wave
vector transfer $\bbox{Q}=(0,0,2.5)$ ($Q_{\|}=5\pi/2$), using
Setup IV (Fig.~\ref{zone-boundary}). The observed peak was
analyzed using a Gaussian profile (solid line in
Fig.~\ref{zone-boundary}) and the zone-boundary energy was
determined: $\hbar\omega_{\rm ZB} =37.9(0.2)\;\mathrm{meV}$. With
nearest-neighbor spin separation along the chain axis equal to
$c/2$, for an $S=1/2$ 1D AF one expects $\hbar\omega_{\rm ZB}=\pi
J/2 $ and $v_{s}=\pi cJ/4$.\cite{Cloizeaux_Pearson} Assuming the
nearest-neighbor model, for ${\rm BaCu_{2}Si_{2}O_{7}}$ we thus
obtain $J=24.1(1)\;\mathrm{meV}$ and $v_{\rm
s}=130.5(5)\;\mathrm{meV\cdot}$\AA, in excellent agreement with
direct spin wave velocity measurements and magnetic susceptibility
data.\cite{Tsukada_BaCu2Si2O7} The agreement suggests that taking
into account only nearest-neighbor interactions is indeed an
adequate description of the spin chains in ${\rm
BaCu_{2}Si_{2}O_{7}}$.

\section{Discussion}\label{discussions}
Most measurements discussed in this paper have been performed
below $4.5\;\mathrm{meV}$, which is the threshold of the
excitation continuum.\cite{Zheludev_BaCu2Si2O7,tobepublished}
Below this energy, the spectrum is expected to be dominated by
single-particle spin wave excitations. The corresponding dynamic
structure factor, and the spin wave dispersion relation in
particular, can be calculated using the chain-Mean Field
(chain-MF) approximation.\cite{Schulz96,Essler_Tsvelik} As will be
discussed below, this theoretical construct is in remarkably good
quantitative agreement with our experimental results on ${\rm
BaCu_{2}Si_{2}O_{7}}$.

\subsection{Model Hamiltonian}
We shall now describe the model Hamiltonian for  ${\rm
BaCu_{2}Si_{2}O_{7}}$ that was used as a starting point in our
data analysis. As mentioned above, a single nearest-neighbor
Heisenberg exchange coupling constant $J$ is sufficient to
describe in-chain interactions. We found that in order to
reproduce the observed dispersion in the perpendicular directions,
three independent inter-chain exchange constants are required. The
${\rm Cu^{2+}}$ chains in ${\rm BaCu_{2}Si_{2}O_{7}}$ form a
rectangular lattice with nearest-neighbor Cu-Cu distances along
the $a$- and $b$-axes of $a/2$ and $b/2$, respectively. We shall
denote the corresponding exchange constants as $J_{x}$ and $J_{
y}$. In addition, we shall include exchange interactions along the
$[1,1,0]$ direction (3rd nearest-neighbor inter-chain Cu-Cu
distance) in our model and label the corresponding coupling
constant as $J_{3}$. The model Heisenberg Hamiltonian is then
written as:
\begin{eqnarray}
 H  & = &\sum_i  J\bbox{S}_{\bbox{r}_i}\bbox{S}_{\bbox{r}_i+\bbox{c}/2}
 +  J_{x}\bbox{S}_{\bbox{r}_i}\bbox{S}_{\bbox{r}_i+\bbox{a}/2}
 +  J_{y}\bbox{S}_{\bbox{r}_i}\bbox{S}_{\bbox{r}_i+\bbox{b}/2}\nonumber\\
 &+&  J_{3}\bbox{S}_{\bbox{r}_i}\bbox{S}_{\bbox{r}_i+\bbox{a}/2+\bbox{b}/2}
 +  J_{3}\bbox{S}_{\bbox{r}_i}\bbox{S}_{\bbox{r}_i+\bbox{a}/2-\bbox{b}/2}
 \label{hamil}.
\end{eqnarray}
Here the sum is taken over all spins in the system and
$\bbox{r}_i$ is the position of spin $i$.\par

Following Schulz \cite{Schulz96} and treating the interchain
coupling in a mean-field treatment, the Hamiltonian transforms
into an effective single-chain Hamiltonian which reads
\begin{equation}
 H  = \sum_i  J\bbox{S}_{i}\bbox{S}_{i+1}-h\sum_i (-1)^i S^z_i
 - 2NJ'm_0^2\, .
\label{eff_hamil}
\end{equation}
The magnetic moment is aligned along the z-axis. Here N is the
number of sites in the chain, $m_0$=$(-1)^i<S_i^z>$ is the
staggered magnetization, and $h$=$-4J'm_0$. The effective MF
inter-chain coupling strength $J'$ for this anisotropic coupling
geometry is given by $J'\equiv \frac{1}{4}(
2|J_{x}|+2|J_{y}|+4|J_{3}|)$.

\subsection{Dynamic structure factor}
Essler $\textit{et. al.}$ have previously derived the SMA dynamic
structure factor for weakly-coupled $S=1/2$ antiferromagnetic
chains in ${\rm KCuF_{3}}$, using the chain-MF/Random Phase
Approximation (chain-MF/RPA).\cite{Essler_Tsvelik} It is
straightforward to adapt their result to the coupling geometry for
the Hamiltonian given in Eq.~\ref{hamil}:
 \begin{eqnarray}
 S^{xx}(\bbox{Q},\omega) &\propto &
  \frac{1}{\omega_{x}(\bbox{Q})}\left[\delta(\omega-\omega_{x}(\bbox{Q}))-
\delta(\omega+\omega_{x}(\bbox{Q})) \right],\label{Sx}\\
 S^{yy}(\bbox{Q},\omega)& \propto &
  \frac{1}{\omega_{y}(\bbox{Q})}\left[\delta(\omega-\omega_{y}(\bbox{Q}))-
\delta(\omega+\omega_{y}(\bbox{Q})) \right],\label{Sy}\\
 \omega_{x,y}^2(\bbox{Q}) & =& \frac{\pi^2}{4}J^2\sin^2(\pi l) +
 \frac{\Delta^2}{J_{y}+2J_{3}-J_{x}} \nonumber \\
& \times & \left(J_{y}+2J_{3}-J_{x}+ J(\bbox{Q})\right)+ D_{
x,y}^2. \label{Eq_dispersion}
\end{eqnarray}
In this formula $S^{xx}(\bbox{Q},\omega)$ and
$S^{yy}(\bbox{Q},\omega)$ are dynamic structure factors for
excitations polarized along the $a$- and $b$-axes, respectively,
and  $\bbox{Q}=(h,k,l)$ is the wave vector transfer. $\Delta$ is
the gap induced in the quantum spin chains by the effective
staggered exchange field in the magnetically ordered state. This
gap corresponds to the excitation energy at a point in reciprocal
space where inter-chain interactions cancel out at the RPA level,
e.g., at  $\bbox{Q}=(0.5,0.5,1)$. $J(\bbox{Q})$ is the Fourier
transform of inter-chain coupling, and is given by
\begin{eqnarray}
&J(\bbox{Q})=J_{x}\cos(\pi h)+J_{y}\cos(\pi k)+ & \nonumber
\\ & J_{3}\cos(\pi(h+k))+J_{3}\cos(\pi(h-k))&.
\end{eqnarray}
Finally, the parameters $D_{x}$ and $D_{y}$ in
Eq.~\ref{Eq_dispersion} are anisotropy gaps for the two spin wave
branches, that we empirically include in the dispersion relation,
as in Tsukada \textit{et al.} \cite{Tsukada_BaCu2Si2O7}.\par

The magnetic dynamic structure factor is related to the inelastic
neutron scattering cross section through the polarization factors
and form factors for the magnetic ions involved:
\begin{eqnarray}
&\frac{d \sigma}{d \Omega d E'}\propto |f(Q)|^2 \left[
S^{xx}(\bbox{Q},\omega)\sin^2(\widehat{\bbox{Q}},\widehat{\bbox{a}})+
\right.&\\&\left.S^{yy}(\bbox{Q},\omega)\sin^2(\widehat{\bbox{Q}},\widehat{\bbox{b}})
\right]\, .& \label{intens}
\end{eqnarray}

The model cross section defined by Eqs.~\ref{Sx}-\ref{intens} were
numerically convoluted with the calculated spectrometer resolution
function and used in global least-squares fits to the data for
each experimental setting. The effective MF inter-chain coupling
strength $|J'|\equiv \frac{1}{4}( 2|J_{x}|+2|J_{y}|+4|J_{3}|)$ is
related to the gap energy through
$\Delta=6.175|J'|$.\cite{Schulz96} The in-chain exchange constant
was fixed at $J=24.1\;\mathrm{meV}$, while the parameters $J_{x}$,
$J_{y}$, $J_{3}$, $D_{x}$ and $D_{y}$ were refined. The following
values were obtained: $J_{x}=-0.460(7)\;\mathrm{meV}$
(ferromagnetic), $J_{y}=0.200(6)\;\mathrm{meV}$,
$2J_{3}=0.152(7)\;\mathrm{meV}$, $D_{x}=0.36(2)\;\mathrm{meV}$ and
$D_{y}=0.21(1)\;\mathrm{meV}$. This yields
$\Delta=2.51\;\mathrm{meV}$ and $|J'|=0.41\;\mathrm{meV}$. The
obtained dispersion relation is shown in Fig.~\ref{dispersion} in
solid lines. Symbols indicate excitation energies obtained in fits
to individual scans.


\subsection{Comparison with the chains-MF model}

The chain MF theory does not only reproduce the observed
dispersion relation, but is in fact in good quantitative agreement
with the data. According to Schulz \cite{Schulz96} the relations
between $m_{0}$, $T_{N}$ and $|J'|$ are given by:
\begin{equation}
|J'|= \frac {T_{N}} {1.28 \sqrt{\ln(5.8J/T_{N})}  }
\end{equation}
\begin{equation}
m_{0}=1.017\sqrt{|J'|/J}\, .
\end{equation}In our case, where $T_{N}=9.0\;\mathrm{K}$
and $J=24.1\;\mathrm{meV}$, this gives $|J'|=0.27\;\mathrm{meV}$
and $m_{0}=0.11\;\mu_{B}$. The remaining small discrepancy with
experiment can be attributed to the presence of weak easy-axis
(Ising-like) anisotropy, which clearly should have the effect of
enhancing long-range order. Indeed, this term, no matter how
small, induces long-range order even in the purely 1D model.

\section{Conclusion}
The obtained detailed characterization of  ${\rm
BaCu_{2}Si_{2}O_{7}}$ provides a quantitative basis for the
discussion of single-particle vs. continuum excitations in
weakly-coupled spin chains in Zheludev \textit{et al.}
\cite{Zheludev_BaCu2Si2O7}. The experimental data are remarkably
well described by the quantum mean field model.

\begin{acknowledgments}
We would like to thank Dr. L.~P. Regnault (CEA Grenoble) and Dr.
A. Wildes for their assistance with experiments at ILL, Dr. S.-H.
Lee for his assistance with preliminary measurements at NIST,
Prof. A. Tsvelik (Oxford University), Prof. R.~A. Cowley (Oxford
University) and Dr. I. Zaliznyak (BNL) for illuminating
discussions, and Mr. R. Rothe (BNL) for technical support. This
work is supported in part by the U.S.-Japan Cooperative Program on
Neutron Scattering, Grant-in-Aid for COE Research ``SCP coupled
system" of the Ministry of Education, Science, Sports, and
Culture. Work at Brookhaven National Laboratory was carried out
under Contract No. DE-AC02-98CH10886, Division of Material
Science, U.S.\ Department of Energy. ORNL is managed for the U.S.
D.O.E. by UT-Battelle, LLC, under contract no. DE-AC05-00OR22725.
One of the authors (M.~K.) is supported by a TMR-fellowship from
the Swiss National Science Foundation under contract no.
83EU-053223.
\end{acknowledgments}


\begin{thebibliography}{20}
\bibliographystyle{prsty}
\bibitem{H_A_Bethe}H.~A. Bethe, Z. Phys. {\bf 71},  205  (1931).
\bibitem{Faddeev_Takhatajan}L.~D. Faddeev and L.~A. Takhatajan,
Phys. Lett. A {\bf 85},  375  (1981).
\bibitem{G_Muller}G. M\"{u}ller, H. Thomas, H. Beck, and J.~C.
Bonner, Phys. Rev. B {\bf 34}, 1429 (1981).
\bibitem{Tsukada_BaCu2Si2O7}I. Tsukada, Y. Sasago, K. Uchinokura,
A. Zheludev, S. Maslov, G. Shirane, K. Kakurai, and E. Ressouche,
Phys. Rev. B {\bf 60},  6601  (1999).
\bibitem{Oliveira_Ba2CuSi2O7} J.~A.~S. Oliveira, Ph.D. thesis,
Ruprecht-Karls-Universit\"{a}t, Heidelberg, 1993.
\bibitem{Satija_KCuF3}S.~K. Satija, J.~D. Axe, G. Shirane,
H. Yoshizawa, and K. Hirakawa, Phys. Rev. B {\bf 21}, 2001 (1980).
\bibitem{Bonner_Fisher}J.~C. Bonner and M.~E. Fisher, Phys. Rev.
{\bf 135}, A640 (1964).
\bibitem{Zheludev_BaCu2Si2O7}A. Zheludev, M. Kenzelmann, S. Raymond,
E. Ressouche, T. Masuda, K. Kakurai, S. Maslov, I. Tsukada, K.
Uchinokura, and A. Wildes, Phys. Rev. Lett. {\bf 85}, 4799 (2000).
\bibitem{D_Tennant_2}D.~A. Tennant, S.~E. Nagler, D. Welz, G. Shirane,
and K. Yamada, Phys. Rev. B {\bf 52},  13381  (1995).
\bibitem{Kojima_Sr2CuO3_Ca2CuO3}K.~M. Kojima, Y. Fudamoto, M. Larkin,
G.~M. Luke, J. Merrin, B. Nachumi, Y.~J. Uemura, N. Motoyama, H.
Eisaki, S. Uchida, K. Yamada, Y. Endoh, S. Hosoya, B.~J.
Sternlieb, and G. Shirane, Phys. Rev. Lett. {\bf 78}, 1787 (1997).
\bibitem{Tsukada_BaCu2Ge2O7}I. Tsukada, J. Takeya, T. Masuda,
and K. Uchinokura, Phys. Rev. B {\bf 62}, R6061
(2000).
\bibitem{Cooper_Nathans}M.~J. Cooper and R. Nathans, Acta
Crys. {\bf 23}, 357 (1967).
\bibitem{Cloizeaux_Pearson}J. des Cloizeaux and J.~J. Pearson, Phys. Rev.
{\bf 128}, 2131 (1962).
\bibitem{tobepublished}A. Zheludev \textit{et al.}, to be published.
\bibitem{Schulz96}H.~J. Schulz, Phys. Rev.
Lett. {\bf 77}, 2790 (1996).
\bibitem{Essler_Tsvelik}F.~H.~L. Essler, A.~M. Tsvelik, and G. Delfino,
Phys. Rev. B {\bf 56}, 11001 (1997).
\bibitem{Yamada_Ca2CuO3}K. Yamada, J. Wada, S. Hosoya, Y. Endoh,
S. Noguchi, S. Kawamata, and K. Okuda, Physica {\bf 253C}, 135
(1995).
\bibitem{Hutchings}M.~T. Hutchings, E.~J. Samuelsen, G. Shirane,
and K. Hirakawa, Phys. Rev. {\bf 188}, 919 (1969).
\end{thebibliography}
\end{document}